\documentclass[preprint,showpacs,prl]{revtex4}
  \usepackage{graphicx}
 \input{epsf}

\begin{document}
 
\title{Evidence for a disorder driven phase transition in 
the condensation of $^4$He in aerogels }

\author{T. Lambert, F. Despetis (*), L. Puech and P.E. Wolf}
\affiliation{Centre de Recherche sur les Tr\`{e}s Basses Temp\'{e}ratures,
laboratoire associ\'e \`a l'Universit\'e Joseph Fourier, C.N.R.S., BP 166,
38042 Grenoble-Cedex 9, France\\
(*)Laboratoire des Collo\"{i}des, Verres, et Nanomat\'eriaux,  CNRS-Universit\'e Montpellier II
Case Courrier 069, 34095 Montpellier-Cedex 5, France \\
}

\date{\today}

\begin{abstract}
We report on thermodynamic and optical measurements of the
condensation process of $^4$He in two silica aerogels of same porosity
95\%, but different microstructures resulting from different pH during
synthesis.  For a base-catalyzed aerogel, the temperature dependence
of the shape of adsorption isotherms and of the morphology of the
condensation process show evidence of a disorder driven transition,
in agreement with recent theoretical predictions.  This transition is
not observed for a neutral-catalyzed aerogel, which we interpret as
due to a larger disorder in this case.
\end{abstract}


\maketitle

Condensation of fluids in mesoporous media, i.e. with pores sizes
falling in the nm-$\mu$m range, has long been the subject of extended
research, both fundamental and applied \cite{Gelb99}.  From the point
of view of statistical physics, this phenomenon offers an experimental
realization of a phase transition in presence of disorder and
confinement.  From the practical point of view, it is widely used as a
way to characterize the pore distribution.  The most common case is
that of sponge-like porous media, where the fluid is confined in
well-defined cavities, either highly connected as, e.g., in Vycor
glass, or with little or no connection, as in MCM41 silica.  In the
case where the dense phase of the fluid wets the substrate, one
observes the progressive adsorption of a thin film at low vapor
pressure, followed by a rather abrupt filling at a pressure of order,
but smaller, than the saturated vapor pressure.  This phenomenon is
coined capillary condensation, being driven by the surface energy
between the liquid and the vapor.  In the classical picture, the
condensation pressure depends on this surface energy and the pore size
through the so-called Kelvin equation \cite{Gelb99}.  The process is
hysteretic, emptying occurring at a lower pressure than filling.
While this hysteresis is well understood at the level of the unique
pore, its origin remains debated in the case of disordered porous
media, with connected pores of different sizes \cite{Gelb99, Knorr04}.
In particular, it has been recently shown by Kierlik \it{et al }\rm
\cite{Kierlik01} that it could result from the disordered nature of
the porous media, rather than, as usually thought, of its geometry or
topology (`pore-blocking').  In this new description, for a given
pressure close to saturation, there exists a large number of
metastable equilibrium states, the one being selected depending on the
time history of the system, which gives rise to the observed
hysteresis.

As such, this description should apply to porous media of very
different topology, silica aerogels\cite{DetcheverryE03}.  In these
gels, the silica forms a complex arrangement of interconnected
strands, resembling more a three-dimensional spider web than a sponge.
Despite the positive curvature of the strands , which unfavors the
film adsorption, aerogels fill below the saturated pressure, which, in
a `classical' picture, has been attributed to capillary condensation
starting on the intersections between strands \cite{Scherer98}.
Experiments are however delicate to interpret as, due to their very
low elastic modulus, the gels deform a lot upon adsorption of fluids
with usual values of surface tension, such as
nitrogen \cite{Reichenauer01}.  This leads to the idea of
studying these gels using Helium, which has a much lower surface
tension.  Such an experiment, albeit differently motivated, was first
performed by Wong and Chan, about 15 years ago \cite{Wong90}.  Their
results showed capillary condensation, but, quite surprisingly, at a
well defined pressure rather than on some range, as one would expect
considering the wide distribution of scales in the silica aerogels.
Together with specific heat experiments, this lead these authors to
propose the existence of a genuine phase transition of helium inside
aerogels, with a slightly depressed critical temperature with respect
to the bulk case ($T_{c}=5.195 K$).  However, latter experiments by
other groups performed on similar aerogels (95\% porosity, gelation
process with basic pH), did not confirm this conclusion.  The sorption
isotherms are hysteretic, and condensation occurs over a finite range
of pressures \cite{Tulimieri99,Gabay00, HermanTout,Lambert04}.  Still,
this range remains unexpectedly narrow, especially at low porosity
 \cite{Tulimieri99,HermanPRB05} or low temperature \cite{Lambert04,
HermanPRB05}.

Both facts are in agreement with theoretical work
\cite{DetcheverryE03,Detcheverry0405} extending Kierlik \it{et al}\rm
's description to a 95\% porosity aerogel, numerically synthesized
using diffusion limited cluster aggregation (DLCA), a process known to
account for the structure factor of base-synthesized aerogels.  These
studies show a disorder-induced transition similar to that predicted
by the Random Field Ising Model at zero temperature \cite{Sethna}.  At
small porosity (or large disorder), filling takes place by a
succession of small, microscopic avalanches, where a limited number of
sites switch from gas to liquid at each pressure.  The maximal size of
these avalanches increases with the porosity, up to some critical
porosity (corresponding to some critical disorder), where it diverges.
Above this critical porosity, which increases with increasing
temperature, filling involves a macroscopic avalanche, associated with
a jump in the average density, at some well defined pressure.  This
implies a change of shape, from smooth to sharp, of the hysteresis
loop, when the porosity is increased at a constant temperature or,
alternatively, when the temperature is decreased below some critical
value, at a constant porosity.  In magnetic systems, an equivalent
change of the hysteresis loop has been observed in Co/CoO disordered
films \cite{Berger00}.  Although the observed evolution of shape with
temperature is an appealing argument in favor of the Kierlik \it{et
al} \rm vision of the condensation in disordered porous media, other
explanations could exist as well.  The aim of this paper is to further
investigate the adequacy of the proposed mechanism, and its possible
dependence on the aerogel structure, using combined optical and
thermodynamic measurements.

We used a one step process \cite{Phalippou} to synthesize two aerogels
of same porosity (95\%), but different microstructures, by changing
the pH from basic to neutral.  Neutron \cite{Vacher} and X-ray
\cite{Wang96} scattering experiments previously performed on similar
samples reveal the sensitivity of the structure to the different
kinetics of the sol-gel process in both cases.  The size of the
building silica units is larger for the base-catalyzed aerogel (B100,
i.e. of density 100 kg/m$^3$) than for the neutral aerogel (N102),
whereas the correlation length is larger for N102 than for B100.  This
results in a fractal structure over a much larger range for N102 than
for B100, and, accordingly, a wider distribution of pores
sizes.  The correlation lengths of our own samples were deduced from
their structure factor $S(q)$ at zero $q$ (as measured by light
scattering) and their known porosities and fractal dimensions, using a
theoretical expression of $S(0)$ which has been experimentally checked
\cite{Frisken95}.  This gives values of order 10~nm for B100 and 20~nm
for N102.  Sorption isotherms were performed on both samples at a
number of temperatures, and, at the same time, the way they scattered
light was studied, so as to gain information on the distribution of
helium inside the aerogels, both at the microscopic and macroscopic
levels. 

The experimental cell is a 4 mm thick copper block, traversed by a
20~mm diameter hole, and closed by two sapphire windows.  Aerogels
samples were sliced from 14 mm diameter cylinders.  In the first
experiment, we used a half disk of B100, about 2.7 mm thick.  In the
second experiment, we studied a 3.8 mm thick full disk of N102.  The
cell is mounted in a cryostat with 8 optical ports 45 $^{\circ}$
apart, and is connected to the room temperature gas-handling system by
two lines, one of which is used to fill or empty the cell through a
regulated flowmeter, and the other is connected to a room temperature
pressure sensor, with a resolution of 0.1~mbar.  The flowrate is
adjusted so that the stage of capillary condensation takes from 15 to
30 hours, which we checked to be long enough to get rate independent
results.

The  helium mass in the aerogel is obtained by integrating the flowrate
over time, and correcting for dead volumes as discussed in ref.\cite{Lambert04}.
It is  converted into a fraction $\Phi$ of dense phase
(`liquid'), assuming that the vapor inside the aerogel has its bulk
value and that the density of the dense phase is that measured
for the fully filled aerogel\cite{Lambert04}.

\begin{figure}[t!]
\includegraphics[width=3.5in]{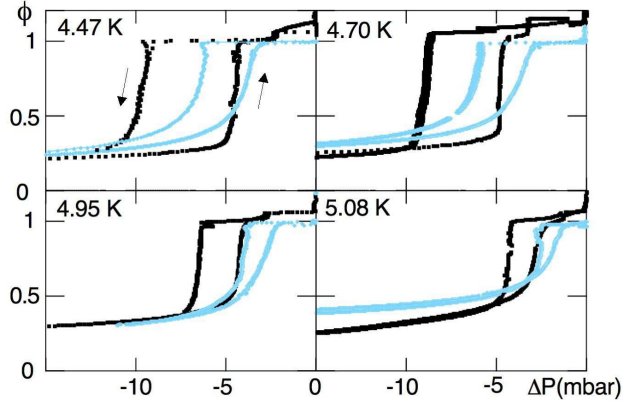}
\caption{(a) Isotherms at four temperatures, 4.47, 4.71, 4.96 and 5.08~K, 
for two aerogels of porosity 95\%, B100 (black) and N102 (grey).  The
liquid fraction $\Phi$ in aerogel is plotted versus $\Delta P
=P-P_{sat}$; for B100, there is a clear change of shape with temperature.}
\label{isothermes}
 \end{figure}

\begin{figure}[h!]
\includegraphics[width=3.5in]{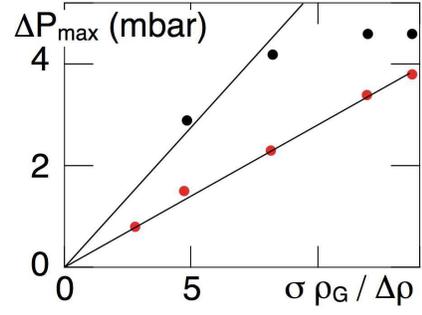}
\caption{Position $\Delta P_{max}$  of the point of maximal slope 
in the adsorption branch versus the temperature dependent ratio 
$\sigma \rho_{V}/(\rho_{L}-\rho_{V})$. The straight lines correspond 
to pore diameters (see text) of 18  and 36~nm. }
\label{positionplateau}
\end{figure}

Figure~\ref{isothermes}(a) shows the isotherms obtained at four
temperatures below the bulk critical temperature $T_{c}$ for the two
aerogel samples.  The general trends are in qualitative agreement with
the usual picture of capillary condensation; the hysteresis loop gets
narrower and closer to the bulk saturation pressure $P_{sat}$ as the
surface tension decreases with increasing temperature.  Also, it is
smoother, and closer to $P_{sat}$ for N102, consistent with the wider
distribution of pores extending up to larger sizes.
Quantitatively, we measure the position of the adsorption isotherm by $\Delta
P_{max}=P-P_{sat}$, at the point of largest slope of the adsorption
branch.  For a cylindrical pore of radius $R$, filling occurs at the
stability limit of the adsorbed film, which, for $R$ much larger than
the film thickness, is given by \cite{Everett,Saam} $\Delta P_{max} =
\sigma/R .  \rho_{V}/(\rho_{L}-\rho_{V})$, where $\sigma$ is the
helium surface tension, and $\rho_{L}$ and $\rho_{V}$ are the bulk
liquid and vapor densities at saturation \cite{note2}.  For a
collection of independent pores, one expects $\Delta P_{max}$ to
correspond to some typical value of the pores size.  Although, \it a
priori \rm, aerogels cannot be so simply described,
figure~\ref{positionplateau} shows that the equation above accounts
for the temperature dependence of $\Delta P_{max}$ for N102 in the
whole temperature range, and for B100 above 4.95~K, with
characteristic `pores' sizes $R\simeq$ 35~nm and 20~nm.  Both values
are about twice the measured correlation lengths, a difference which
is not surprizing, considering the complex microstructure of the
samples.  However, this simple approach fails to describe the position
of the pressure plateau for B100 at the two lower temperatures, which
occurs at a smaller $\Delta P_{max}$ than expected.  Furthermore, in
such an approach, the small width of the corresponding isotherms would
correspond to a much narrower distribution of pores sizes than that
could be deduced from the isotherms at 4.95~K and above.

\begin{figure}[t!]
\includegraphics[width=3.5in]{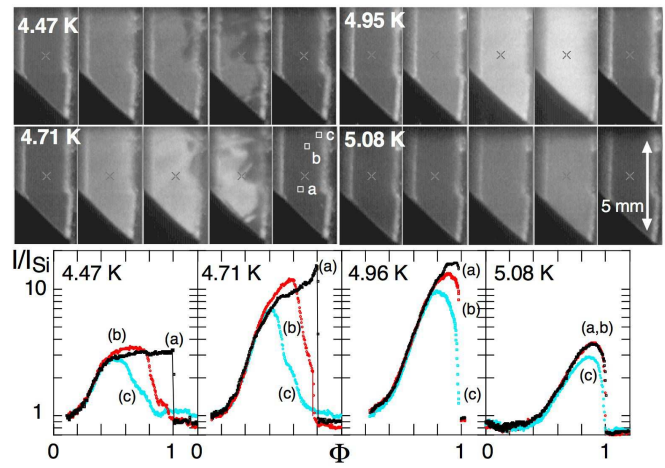}
\caption{Images of B100 observed at 45$^\circ$ from the incident 
laser sheet for helium fractions of 0.1,0.5,0.7,0.9, and 1.0. The 
curves give the scattered intensity, referred to the empty aerogel 
situation, for the three regions labelled (a), (b), (c).}
\label{imageB}
\end{figure}

Optical measurements also show a markedly different behavior below
4.71~K, and above 4.95~K, respectively.  We illuminate the aerogel by
a thin (100 $\mu$m wide), vertically polarized, He-Ne laser sheet
under a 45$^{\circ}$ incidence with respect to its faces, and image it
at 45$^\circ$, 90$^\circ$, and (for N102) 135$^\circ$ using CCD
cameras.  At any point of the intercept of the aerogel by the laser
sheet, the brightness of the image is proportional to the intensity
scattered by the aerogel in the direction of observation, and depends
on the local strength and scale of the spatial fluctuations of the
helium and silica densities.  Figure~\ref{imageB} shows (a) images of
B100 for increasing helium fractions along the previous isotherms, and
(b) the scattered intensity by three selected spots in the images.
Quantitative analysis of these curves show that, below
$\Phi\approx$0.3, the scattered intensity is that expected for a thin
film of helium covering uniformly the silica strands, whereas, beyond
this value, it becomes much larger, implying the formation of liquid
clusters correlated over distances larger than the correlation length
of the silica\cite{Lambert04}.  The striking fact is that, while this
increase of the scattered intensity occurs uniformly over the sample
for the two larger temperatures, for the two lower ones, it is only so
for fractions less than about 0.5.  Beyond, liquid progressively
invades the aerogel, resulting in the growth of dark regions.
Comparison of figures \ref{isothermes} and \ref{imageB}(b) shows that
this invasion occurs along the vertical part of the isotherm.  This
could suggest that the bright and dark regions correspond to two
coexisting phases at thermodynamic equilibrium.  However, this is
ruled out by the fact that the intensity scattered by the bright
regions evolves with $\Phi$, as shown by figure~\ref{imageB}(b) at
4.71~K. On the other hand, the observations exhibit the behavior
expected for macroscopic avalanches ; at any given spot, the
transition to the dark state occurs discontinuously, when this spot is
swept by the boundary between the bright and the dark regions.  This
is consistent with the occurrence of a macroscopic (but local)
avalanche from a partly filled to a fully filled state.  In this
scenario, the macroscopic morphology, which is identical for 4.47~K
and 4.71~K, and for different rates of condensation as well, must
originate in macroscopic, heterogeneities of the average density of
silica, which make some regions favor slightly more the liquid.  This
will make the pressure at which the avalanche takes place to vary
continuously in space, so that the transition should occur
successively at different locations in the aerogel, in qualitative
agreement with the observations.

\begin{figure}[t]
\includegraphics[width=3.5in]{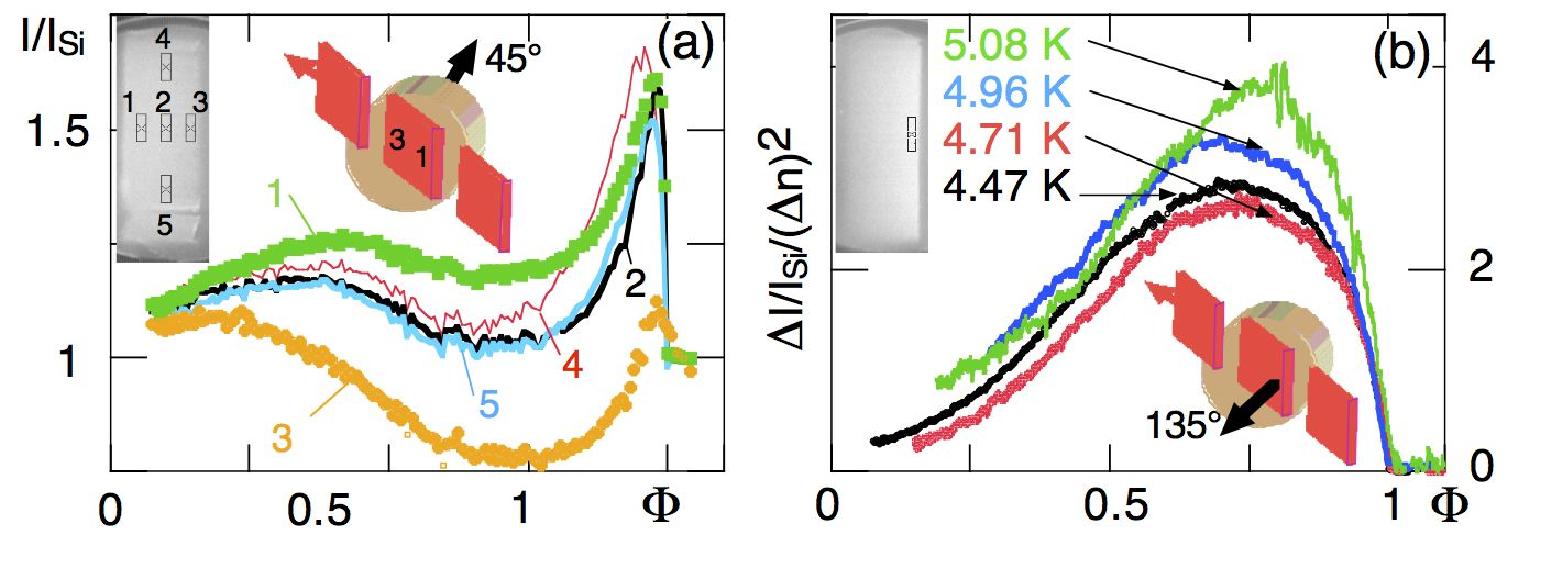}
\caption{ Intensity scattered by N102 as a function of the condensed
fraction $\Phi$ : (a) at 4.47~K and 45$^\circ$.  The signal,
normalized by the silica contribution, corresponds to five regions at
three different depths inside the aerogel disk.  Differences between
curves for different depths (and path lengths) are consistent with
multiple scattering effects combined with an homogeneous distribution
of the liquid clusters within the sample, in contrast to B100 at the
same temperature; (b) at 135$^\circ$, close to the entrance of the
laser sheet.  The silica contribution $I_{Si}$ has been subtracted,
and the result normalized by $I_{Si}$ and the optical contrast
squared.  The signal is similar at all temperatures, suggesting that
the spatial organization of the liquid only depends on $\Phi$, not on
temperature.  }
\label{imageN}
\end{figure}

In summary, for B100 at 4.71~K and below, the isotherms exhibit a
nearly vertical portion, their position is anomalous, and a
significant fraction of the condensation takes place in an
heterogeneous way.  These facts are consistent with the occurrence of a
disorder induced transition for B100 between 4.95~K and 4.71~K.
In contrast, for N102 at all temperatures,
the isotherms are smooth, and the scattering is homogeneous, as shown
in figure~\ref{imageN}(a)~:~as a function of the condensed fraction
$\Phi$, the intensity scattered at 45$^\circ$ is the same along any
vertical line of the picture, corresponding to a given depth, hence
path length, inside the sample.  There is a difference between points
at different depths, but this (as well as the bump in the signal at
$\Phi \approx$ 0.5) can be explained by multiple scattering effects,
which are stronger for N102 than for B100, due to the larger thickness
and silica correlation length.  Hence, no transition is observed for
N102 in the temperature range where it is for B100.  In the
disorder induced transition scenario, this implies that the disorder
does not only depend on porosity and is larger for N102 than for B100.  We
speculate that this is due to the wider
distribution of pores sizes in this case.  It would be interesting to
see whether calculations performed on 95\% aerogels numerically
synthesized using the Diffusion Limited Aggregation algorithm
appropriate for neutral aerogels would indeed confirm the absence of
transition in this case.

Finally, as shown by figure~\ref{imageN} 
(b), the intensity scattered at 135$^\circ$ close to the entrance of 
the laser sheet, once corrected by the contrast factor( $(\Delta 
n)^{2}$, where $\Delta n$ is the difference in refractive index 
between the liquid and the vapor) behaves roughly the same, whatever 
the temperature. This suggests that, at a given $\Phi$, the 
distribution of liquid does not depend on temperature. This is not 
the case for B100 below 4.95~K. As shown by figures~\ref{imageB}(a) 
and (b), the scattered intensity, at a given $\Phi$, is smaller at  
4.46~K than at 4.71~K, although the contrast is larger. This implies 
that the liquid must be distributed in smaller clusters. Here also, 
theoretical calculations would be needed in order to see whether such 
a temperature dependence follows from the disorder induced transition scenario.

In conclusion, our results provide a strong evidence for the occurrence
of a disorder induced transition in the condensation of $^{4}$He in
base-catalyzed aerogels.  Planned experiments using lighter aerogels
should allow to check whether the transition moves to larger
temperatures, as the theory would predict.  
Here again, the very low index of refraction of helium, which allows
quantitative single scattering measurements to be performed over a
much wider range of conditions than for any other simple fluid, will
be a unique advantage.  This ability to perform optical studies in
transparent porous media could be used to study other problems as
well, such as the origin of the hysteresis between adsorption and 
desorption.

We acknowledge T. Herman, J. Beamish, F. Detcheverry, 
E. Kierlik, M.L. Rosinberg, and J. Phalippou for useful 
suggestions or discussions.

\begin{thebibliography}{999}
\bibitem{note2}
{This $\Delta P_{max}$ is half the value given by the Kelvin equation, which
corresponds to the equilibrium of an hemispherical meniscus in the
pore.}
\end{thebibliography}
\end{document}